# Two-Way One-Counter Nets Revisited

**Shaull Almagor**
Department of Computer Science, Technion, Israel

**Michaël Cadilhac**
DePaul University, USA

**Asaf Yeshurun**
Department of Computer Science, Technion, Israel

## Abstract

One Counter Nets (OCNs) are finite-state automata equipped with a counter that cannot become negative, but cannot be explicitly tested for zero. Their close connection to various other models (e.g., PDAs, Vector Addition Systems, and Counter Automata) make them an attractive modeling tool.

The two-way variant of OCNs (2-OCNs) was introduced in the 1980's and shown to be more expressive than OCNs, so much so that the emptiness problem is undecidable already in the deterministic model (2-DOCNs).

In a first part, we study the emptiness problem of natural restrictions of 2-OCNs, under the light of modern results about Vector Addition System with States (VASS). We show that emptiness is decidable for 2-OCNs over *bounded languages* (i.e., languages contained in $a_1^* a_2^* \cdots a_k^*$), and decidable and Ackermann-complete for *sweeping* 2-OCNs, where the head direction only changes at the end-markers. Both decidability results revolve around reducing the problem to VASS reachability, but they rely on strikingly different approaches. In a second part, we study the expressive power of 2-OCNs, showing an array of connections between bounded languages, sweeping 2-OCNs, and semilinear languages. Most noteworthy among these connections, is that the bounded languages recognized by sweeping 2-OCNs are precisely those that are semilinear. Finally, we establish an intricate pumping lemma for 2-DOCNs and use it to show that there are OCN languages that are not 2-DOCN recognizable, improving on the known result that there are such 2-OCN languages.

**2012 ACM Subject Classification** Theory of computation → Automata extensions

**Digital Object Identifier** 10.4230/LIPIcs.CSL.2025.19

**Related Version** *Full Version*: https://arxiv.org/abs/2410.22845

**Funding** *Shaull Almagor*: supported by the ISRAEL SCIENCE FOUNDATION (grant No. 989/22)

## 1 Introduction

A *One-Counter Net (OCN)* is a finite state automaton equipped with a counter that cannot decrease below zero, but cannot be explicitly tested for zero. It is a natural restriction of several computational models: One-Counter Automata without zero tests, and Pushdown Automata with a single letter stack alphabet. It can also be thought of as 1-dimensional Vector Addition Systems with accepting States and an alphabet.

OCNs are an attractive model for studying the border of decidability, as several problems for them lie close to the decidability frontier (e.g., both determinization and universality may be decidable or undecidable, depending on the precise definition and context [16, 17, 2, 3, 1]).

OCNs suffer from an unsettling asymmetry: the language $L = \{a^n b^m \mid n \geq m\}$ is OCN-recognizable (by counting $+1$ on $a$ and $-1$ on $b$), whereas the language $\{a^n b^m \mid n \leq m\}$ is not OCN-recognizable. In particular, they are not closed under reversal (nor under intersection). A natural way of making OCNs more robust is therefore to look at their two-way variant.

This approach was taken in [26, 7], where the model of *two-way one-counter nets (2-OCNs)* is introduced and studies. A 2-OCN is a one counter net that receives its input on a tape, surrounded by end-markers $\vdash$ and $\dashv$, and is allowed to move a read-only head back and forth on the tape. As the following examples witness, the introduction of a two-way tape significantly increases the expressive power of the model, as well as its deterministic fragment *(2-DOCN)*. For uniformity, we henceforth refer to one-way OCNs as *1-OCN*.

▶ **Example 1.** Consider the language $L_1 = \{a^n b^n \mid n \in \mathbb{N}\}$, which is easily shown not to be 1-OCN-recognizable. $L_1$ can be recognized by a 2-OCN (in fact 2-DOCN, see Fig. 1) by using a "counter reset": it starts with a forward scan verifying that the format is $a^n b^m$ while counting $+1$ on $a$ and $-1$ on $b$. Thus, upon reaching $\dashv$, the counter has value $n - m$ (unless $n < m$, in which case the run terminates due to the counter becoming negative, and the word is not accepted), and in particular $n \geq m$. It then moves backwards to $\vdash$, counting $-1$ on $a$ and $+1$ on $b$, thus resetting the counter back to 0. Next, it goes forward again to $\dashv$, then similarly computes $m - n$ from right to left. Intuitively, this approach recognizes $\{a^n b^m \mid n \geq m\} \cap \{a^n b^m \mid n \leq m\}$ and uses the closure of 2-DOCN under intersection [26].

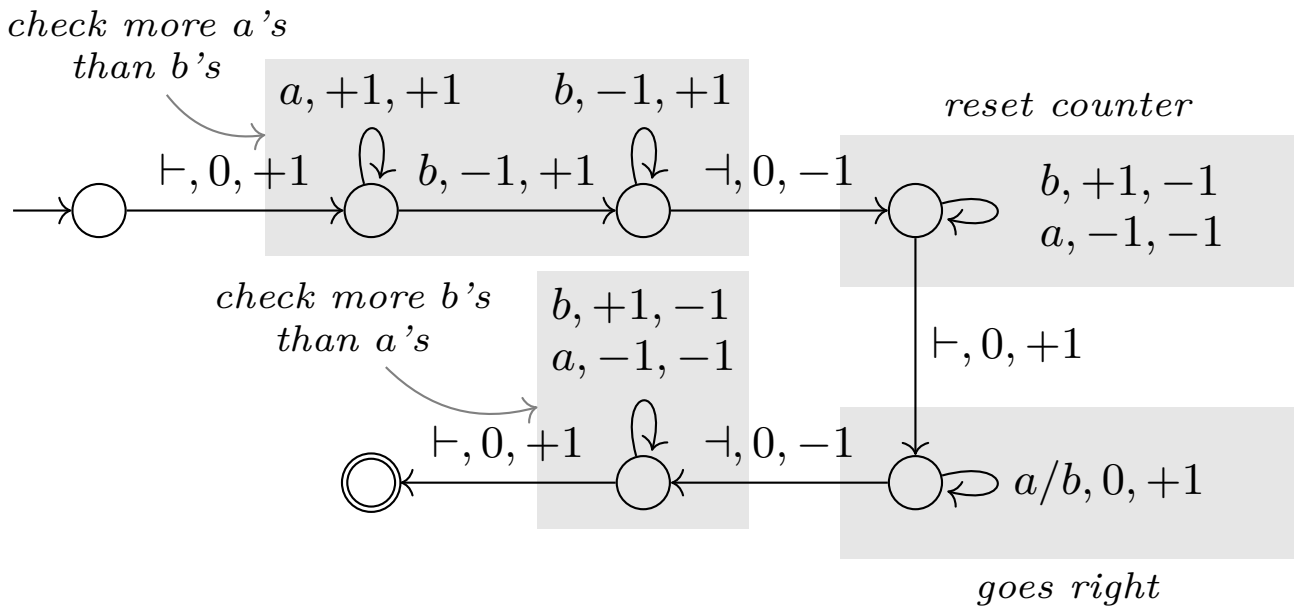

■ **Figure 1** 2-DOCN recognizing $\{a^n b^n \mid n \in \mathbb{N}\}$ using counter-reset. A transition $\sigma, e, h$ means "read letter $\sigma$, change counter by $e$, and move head by $h$".

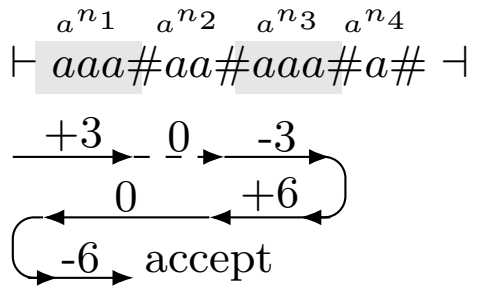

■ **Figure 2** An accepting run of a 2-OCN for $L_2$. The dashed line signifies a nondeterministic guess.

The next example demonstrates the fact [7] that 2-OCN are more expressive than 2-DOCN. In addition, it demonstrates that sometimes it is (seemingly) necessary for the head to "change direction" inside the tape, rather than at the end markers.

▶ **Example 2.** Let $L_2$ be the language $\{a^{n_1}\# \cdots \# a^{n_k}\# \mid k \in \mathbb{N} \wedge (\exists i > 1)[n_1 = n_i]\}$, i.e., words where the length of the first $a$-segment is matched in some later segment. This language is not 1-OCN recognizable, and one may observe that a similar approach to Example 1 fails, because we would need to guess the index $i$ for which $n_1 = n_i$, but in order to reset the counter, we must "forget" the guess. Thus, in order to recognize this with a 2-OCN, we must compare $n_1$ to $n_i$ via a single visit to the $i$-th segment.

To achieve this, the 2-OCN counts $+1$ on the first segment $a^{n_1}$, then scans forward and guesses when the correct segment $i$ is reached. It then counts $-1$ on $a^{n_i}$. Thus, upon reaching $\#$, the counter holds $n_1 - n_i$ (and so guarantees that $n_1 \geq n_i$ if it survives). Then, a backward scan is performed from $\#$ counting $+2$ on $a$ in $a^{n_i}$, until $\#$ is reached, and then continues left to $\vdash$ and finally counts $-2$ on the $a$'s in $a^{n_1}$. Thus, the counter ends with value $n_1 - n_i + 2n_i - 2n_1 = n_i - n_1$. Again, if $n_i < n_1$ the run terminates, so the whole run survives if $n_1 \leq n_i$, which combined with $n_1 \geq n_i$ gives $n_1 = n_i$ (see Fig. 2).

A 2-OCN that only changes its head direction at the end marks is called *sweeping*. Example 2 prompts studying the expressive power of sweeping 2-OCNs, which we take on in this work.

**Related Work.** Two-way automata have received much attention since their introduction for finite automata in [30]. 2-PDAs were studied in [24, 14, 4, 27], focusing mainly on closure properties and languages recognizable by 2-PDAs. As noted in [4], languages not expressible by 2-PDA can be derived from the complexity results obtained in [14] (specifically, 2-PDA languages are in linear space and polynomial time, and the time/space hierarchy theorems offer languages outside of these classes). No automata-based, combinatorial techniques are known to the authors to show that some languages is not expressible with a 2-PDA, and most questions about 2-PDA, including whether the deterministic variant can recognize all context-free languages, are open [13].

In the world of counter machines, [12] shows that DFAs with a “blind tape” (a form of a nonnegative counter that cannot be 0-tested) are less expressive than 2-DPDAs. A significant body of work by Ibarra et al. studies *reversal bounded* counters [19, 21, 18, 10, 20], i.e., counters that cannot increase and decrease unboundedly many times. The heavy reliance on reversal-boundedness makes these works orthogonal to ours. Atig and Ganty [5] study a VASS-CFG hybrid, and use also make use of reductions to VASS with 0-tests, but with significant technical differences.

Closer to our setting are *two-way one-counter automata* (2-OCAs), that allow explicit zero tests. In [29] certain languages computing squaring or exponentiation are shown not to be 2-DOCA recognizable (the deterministic variant), whereas [11] exhibit nonsemilinear 2-OCA languages. Ibarra and Su [21] note that 1-sweep 2-DOCA have an undecidable emptiness problem; in contrast, we show that this problem is decidable for sweeping 2-OCN. They also note that, over bounded languages, emptiness of 2-DOCA is decidable *provided* that the counter reaches zero only a constant number of times; in contrast, we show that this emptiness over bounded languages is decidable for any 2-OCN.

A nearly identical model to 2-OCN has been introduced in [26] and further studied in [7]. In their model, the counter updates are $\pm 1$, which essentially amounts to encoding counters in unary. In [26], the focus is on 2-DOCN recognizable languages. It is shown that this class is closed under intersection, but not under complementation nor union. Further, over bounded languages, 2-DOCN can be made sweeping[1], and the Parikh images of their languages are semilinear. An example is also given to show that 2-DOCN are less expressive than 2-OCN.

The algorithmic contribution in [7] is that the emptiness problem for 2-DOCN is undecidable. Additionally, it is shown that if a 2-OCN accepts a word of length $n$, then it accepts it with a run of polynomial length in $n$. The latter result no longer holds for our model, due to the binary encoding of the counters.

**Contribution and Paper Organization.** In this work we provide a modern view of 2-OCNs and some of its interesting and decidable fragments, namely *bounded-language 2-OCNs* and *sweeping 2-OCNs*. In Section 3.1 we give a counter-machine based proof of the undecidability of 2-DOCN emptiness. In Section 3.2, we consider 2-OCNs whose languages are *bounded*, i.e., of the form $L \subseteq a_1^* a_2^* \cdots a_k^*$. We show that under this restriction, emptiness becomes decidable. In Section 3.3, we consider *sweeping* 2-OCNs, i.e., 2-OCNs whose head-movement direction only changes at the end-markers. We show that here too emptiness becomes decidable and we give precise complexity bounds, using recent results regarding VASS reachability. In Section 4, we investigate the relationship between semilinear languages and 2-OCN languages. We show that for bounded languages, the set of *lengths* of words in the language of a 2-OCN is semilinear, and derive explicit languages that are not 2-OCN-recognizable.

[1] The proof in [26] is actually missing significant details, but those can be reconstructed.

We also show that a bounded language is semilinear if and only if it is recognizable by a sweeping 2-OCN. In Section 5, we show an elaborate pumping lemma for 2-DOCNs. Apart from its independent usefulness for understanding 2-DOCNs, this enables us to demonstrate a 1-OCN whose language is not 2-DOCN recognizable, establishing a new separation result. We conclude in Section 6 with some open problems.

We encourage the reader to initially skip the proofs and garner a bird's eye view of our results. Thereafter, we note that Section 3.2, Lemmas 15 and 17, and Theorem 14 have the most interesting proofs. Detailed proofs appear in the appendix.

## 2 Preliminaries

**Two-Way OCN and Automata.** For an alphabet set $\Sigma$, we denote $\Sigma_{\vdash\dashv} = \Sigma \cup \{\vdash, \dashv\}$ where $\vdash, \dashv$ are two symbols not appearing in $\Sigma$, called *left end-marker* and *right end-marker*, respectively. A *Two-Way One-Counter Net (2-OCN)* is a tuple $\mathcal{A} = \langle Q, \Sigma, \Delta, q_{\text{init}}, q_{\text{acc}} \rangle$ where $Q$ is a finite set of *states*, $\Sigma$ is a finite alphabet, $\Delta \subseteq Q \times \Sigma_{\vdash\dashv} \times \mathbb{Z} \times \{-1, +1\} \times Q$ is a finite set of *transitions*, $q_{\text{init}}, q_{\text{acc}} \in Q$ are the initial and accepting states, respectively. Intuitively, a transition $\tau = (q, \sigma, e, h, q')$ dictates that from state $q$, when reading letter $\sigma$, we add $e$ to the counter value, move the position of the head on the tape by $h$, and reach state $q'$. We call $e$ the *effect* and $h$ the *head shift* of the transition. Unless explicitly stated otherwise, we assume that the effect $e$ is encoded in binary. We require that if $(q, \vdash, e, h, q') \in \Delta$ then $h = 1$ and if $(q, \dashv, e, h, q') \in \Delta$ then $h = -1$. That is, upon reaching the end-markers, the head cannot proceed beyond them.

We say that $\mathcal{A}$ is *deterministic* (denoted 2-DOCN) if for every $q \in Q, \sigma \in \Sigma_{\vdash,\dashv}$ there is at most one $q' \in Q$ and $e, h$ such that $(q, \sigma, e, h, q')$. We say that $\mathcal{A}$ is *one way* (denoted 1-OCN) if every transition moves the head to the right (i.e., the head shift is 1).

A *configuration* of $\mathcal{A}$ is $(q, c, p) \in Q \times \mathbb{N} \times \mathbb{N}$ representing the current state $q$, counter value $c$, and head position $p$. Note that we do not consider configurations with negative counter values, thus enforcing the semantics of Counter Nets. Consider a word $w = \sigma_1 \cdots \sigma_m \in \Sigma^*$, we augment it to $\vdash w \dashv$ by setting $\sigma_0 = \vdash$ and $\sigma_{m+1} = \dashv$. A *run* of $\mathcal{A}$ on $w$ is a finite sequence of configurations $\rho = (q_1, c_1, p_1), (q_2, c_2, p_2), \ldots, (q_n, c_n, p_n)$ such that $0 \leq p_i \leq m+1$ for all $i$, and for every $1 \leq i < n$ we have $(q_i, \sigma_{p_i}, c_{i+1} - c_i, p_{i+1} - p_i, q_{i+1}) \in \Delta$. That is, each configuration follows from the previous one by the transition relation. A run is *initial* if $q_1 = q_{\text{init}}$ and $p_1 = c_1 = 0$. A run is *accepting* if $q_n = q_{\text{acc}}$. We tacitly assume that runs do not continue after the accepting state. The word $w$ is accepted by $\mathcal{A}$ if $\mathcal{A}$ has an initial and accepting run on it. We define $L(\mathcal{A}) = \{w \in \Sigma^* \mid \mathcal{A} \text{ accepts } w\}$. Note that, just like OCNs, 2-OCNs are *monotone*, i.e., every word accepted from initial configuration $(q, c, p)$ is also accepted from $(q, c', p)$ for every $c' > c$. We sometimes consider sequences of pseudo-configurations where the counter value becomes negative. In such cases, we say that the run is *cut short by a counter violation*, and we mean that the sequence of configurations is a run up to a prefix in which the counter becomes negative.

Finally, we sometimes discuss classical Boolean automata (NFA, DFA, 2-NFA and 2-DFA). For our purposes, they can be thought of as 2-OCN where all the counter updates are 0, and are therefore omitted.

**VASS and VASS with bounded 0-tests.** A *Vector Addition System with States (VASS)* is a tuple $\mathcal{V} = \langle S, T \rangle$ where $S$ is a finite set of states and $T \subseteq S \times \mathbb{Z}^d \times S$ is the transition relation. A *configuration* of $T$ is $(s, \boldsymbol{v})$ where $s \in S$ and $\boldsymbol{v} \in \mathbb{N}^d$ (note that this is a vector of nonnegative integers). We say that configuration $(s_2, \boldsymbol{y})$ is *reachable* from configuration $(s_1, \boldsymbol{x})$ if there is a finite sequence of configurations $(p_1, \boldsymbol{v_1}), \ldots, (p_m, \boldsymbol{v_m})$ such that $(p_1, \boldsymbol{v_1}) = (s_2, \boldsymbol{y}), (p_m, \boldsymbol{v_m}) = (s_2, \boldsymbol{y})$ and for every $1 \leq i < m$ we have $(p_i, \boldsymbol{v_{i+1}} - \boldsymbol{v_i}, p_{i+1}) \in T$. Intuitively, a VASS runs by repeatedly adding vectors to the configuration, as long as all the *counters* (i.e., the entries of the vector) remain nonnegative. The reachability problem for VASS asks whether $(s_2, \boldsymbol{y})$ is reachable from $(s_1, \boldsymbol{x})$ and is known to be decidable [9].

While VASS do not allow explicit 0-tests, it is possible to simulate a fixed number of 0-tests within the context of reachability [8, 9]. That is, we can extend the transition relation so that $T \subseteq S \times \mathbb{Z}^d \times \{=0, \geq 0\}^d \times S$, and a transition can be taken only when the counters that have $= 0$ tests are 0. As long as there is a constant $B$ such that there are no more than $B$ 0-tests along every run of the VASS, then reachability remains decidable. The simplest approach to achieve this is to increase the dimension by $d \cdot B$, keeping $B$ copies of every counter, and whenever a counter needs to be tested for 0, a copy of it is "frozen" (i.e., never changes again) and the reachability target has 0 on all frozen entries. Note that we can mark which entries are frozen in the states of the VASS, since there are at most $B$ of them. We remark that more elaborate approaches can keep the number of added counters small [8, 9]. We call such machines *VASS with bounded 0-tests*.

## 3 The Emptiness and Membership Problems in 2-OCNs and Variants

### 3.1 Emptiness of 2-DOCNs is Undecidable

It is shown in [7] that already for 2-DOCNs, the emptiness problem, namely the problem of deciding whether $L(\mathcal{D}) = \emptyset$ for a given 2-DOCN $\mathcal{D}$, is undecidable. This is shown by a reduction from Hilbert's 10th problem. More precisely, it is shown that 2-DOCN can simulate multiplication, in a sense.

An arguably cleaner way of obtaining this result while staying in the world of counter machines, is via reduction from the halting problem for Two-Counter Machines, as follows:

▶ **Theorem 3** ([7]). *The emptiness problem for 2-DOCN is undecidable.*

**Proof.** We show the undecidability of 2-DOCN emptiness by reduction from the (complement of the) halting problem for two-counter machines, which is known to be undecidable [25].

Given a two-counter machine $\mathcal{M}$, we construct a 2-DOCA $\mathcal{A}$ which reads input words of the form $\#(x^*y^*\#)^*$, representing the values of the counters $x$ and $y$ along the run of $\mathcal{M}$, with each step of the machine separated with #. Intuitively, $\mathcal{A}$ tracks the location of $\mathcal{M}$ in its state, uses the counter values to determine the location after a jump, and at each steps checks that the counter values between steps are consistent, similarly to Example 1. Then, $\mathcal{A}$ accepts if at any point the computation reaches the location `halt`. If a counter violation is encountered, $\mathcal{A}$ moves to a rejecting sink.

More precisely, $\mathcal{A}$ starts by checking (using a single right-to-left pass with a DFA) that the word adheres to the format $\#(x^*y^*\#)^*$. It then resets the head to $\vdash$, records the location $l_1$ in its state, moves the head to the first # and begins the simulation. $\mathcal{A}$ simulates each step starting from the # preceding the segment $x^m y^k$ representing the current counter values, and along the simulation scans only the segment of the word of the form $\#x^m y^k \# x^{m'} y^{m'} \#$ starting at the current head location. Note that crucially, the format of the segment is fixed, so that the head can end the simulation in the middle #, thus correctly viewing the counter values of the next step.

If the command at the current location is `goto` $l_i$, then $\mathcal{A}$ checks that $m = m'$ and $k = k'$, similarly to Example 1. Similarly, if the current command is `inc(x)` or `dec(y)`, then $\mathcal{A}$ checks that $k' = k + 1$ and $m = m'$, or that $k = k'$ and $m' = m - 1$, respectively, by a similar comparison prefixed by a $\pm 1$ counter change.

If the current command is `if x=0 goto` $l_i$ `else goto` $l_j$, then $\mathcal{A}$ checks that $k = k'$ and $m = m'$, and then checks whether $k = 0$ (by checking if there are any $x$'s after the first # in the segment), and if so moves to $l_i$ and otherwise to $l_j$.

Finally, if $\mathcal{A}$ ever reaches the command `halt`, it accepts, and if any of the comparisons fails, the counter goes below 0 and the word is rejected. If $\dashv$ is reached without reaching `halt`, then $\mathcal{A}$ moves to a rejecting sink.

We then have that $L(\mathcal{A}) \neq \emptyset$ if and only if there exists a word representing the counter values in a halting run of $\mathcal{M}$, if and only if $\mathcal{M}$ halts. ◀

## 3.2 Emptiness of 2-OCNs over Bounded Languages is Decidable

Consider an alphabet $\Sigma = \{a_1, \ldots, a_k\}$. A language $L$ is *bounded* if $L \subseteq a_1^* a_2^* \cdots a_k^*$. In this section, we show that the emptiness problem of 2-OCNs over bounded languages is decidable.

▶ Remark 4. Generally, a bounded language is a subset of $x_1^* x_2^* \cdots x_k^*$ where each $x_i \in \Sigma^*$ (in particular, e.g., $a^* b^* a^*$ is also bounded). Our results readily extend to these languages, but they add a cumbersome notational layer, which we opt to avoid.

▶ **Theorem 5.** *The following problem is decidable: given a 2-OCN $\mathcal{A}$ is $L(\mathcal{A}) \cap a_1^* a_2^* \cdots a_k^* = \emptyset$?*

We prove Theorem 5 in the remainder of this section. The proof is by reduction to the VASS nonreachability problem, known to be Ackermann-complete [9].

Consider a 2-OCN $\mathcal{A}$. We henceforth assume that membership in $a_1^* a_2^* \cdots a_k^*$ is already tested within $\mathcal{A}$, i.e., that $L(\mathcal{A})$ is a bounded language. We further assume that $L(\mathcal{A}) \subseteq a_1^+ a_2^+ \cdots a_k^+$, namely that every letter occurs at least once in every word in $L$ (dubbed *properly-bounded*). Indeed, for the purpose of emptiness we can split $L(\mathcal{A})$ to a union over all subsets $\Gamma \subseteq \Sigma$ of properly-bounded languages over $\Gamma$.

We obtain from $\mathcal{A}$ a VASS with bounded 0-tests $\mathcal{V}$ such that $L(\mathcal{A}) \neq \emptyset$ iff $(s_{\text{init}}, \mathbf{0})$ reaches $(s_{\text{fin}}, \mathbf{0})$ for certain states $s_{\text{init}}, s_{\text{fin}}$ of $\mathcal{V}$.

### Construction of $\mathcal{V}$ – Intuitive Overview

We present the intuition behind $\mathcal{V}$. The formal construction can be found in Appendix A. $\mathcal{V}$ simulates the behavior of $\mathcal{A}$ in several components using several counters. We start by describing the counters of $\mathcal{V}$ and their intuitive roles.

- `c` – tracking the counter of $\mathcal{A}$
- `c_freeze_1` – a copy of `c`, to be frozen at the beginning of a positive loop.
- `c_freeze_2` – a copy of `c`, to be frozen at the end of a positive loop.
- `L[1]` ... `L[k]` and `R[1]` ... `R[k]` – pairs of counters marking the position of the head from the Left and from the Right for each segment $a_i^*$.

The operation of $\mathcal{V}$ is as follows. The *initialization component* guesses a word $a_1^{n_1} \cdots a_k^{n_k}$ by guessing the numbers $n_i$ and storing them in the `R[i]` counters of $\mathcal{V}$. This is done with self loops that allow to arbitrarily increase each of the relevant counters.

Once initialization is complete, $\mathcal{V}$ moves to the *simulation component.* There, $\mathcal{V}$ aims to simulate runs of $\mathcal{A}$ on the guessed word. Simulating the counter of $\mathcal{A}$ is straightforward, using the counter `c` to match it. In addition, if at any point the accepting state of $\mathcal{A}$ is reached, then $\mathcal{V}$ moves to the *finalizing component* (described below). In order to simulate the head movement, we refer to each $a_i^{n_i}$ as a *segment* (treating $\vdash$ and $\dashv$ also as segments), and we split the simulation to two types of moves:

- A *move within a segment* simulates $\mathcal{A}$ reading $a_i$ and the head staying within the $a_i$ segment. To do this, we set the two counters for the $a_i$ segment `L[i]`, `R[i]` such that `L[i]` keeps track of the head location within the segment $a_i^{n_i}$ and `R[i]` $= n_i -$ `L[i]` (using the initialization counter). To simulate a move to the right we increase `L[i]` by 1 and decrease `R[i]` by 1, and vice-versa for a move left. Note that since $n_i$ is constant after initialization, the VASS semantics prevents the simulation from going over the segment borders.
- A *move between segments* occurs when $\mathcal{V}$ guesses that a border of the $a_i$ segment is reached, and an adjacent segment should be moved to. If the move is to the right, then `R[i]` should be 0, and we therefore perform a 0-test on it, and move to segment $a_{i+1}$. Note that from then on, the head movement counters are `L[i+1]` and `R[i+1]` (until $a_i$ is reached again). This segment index is stored in the states of $\mathcal{V}$.
  Since we are only allowed a fixed number of 0-tests, we store in the states how many moves from segment $a_i$ to $a_{i+1}$ were performed. If this number exceeds $|Q| + 1$, where $|Q|$ is the number of states of $\mathcal{A}$, then the run cannot proceed (it reaches a sink). We justify this in the proof.

An accepting run of $\mathcal{A}$ might require the counter to be "charged" by repeating a loop, possibly involving moves between segments. Since we cannot simulate this with a bounded number of 0-tests, we instead detect loops, as follows. While moving between segments, $\mathcal{V}$ may guess that the current move leads to a repetition of the state and head position in the run of $\mathcal{V}$. Thus, $\mathcal{V}$ may nondeterministically record in its state the current state $q$ of $\mathcal{A}$ (i.e., the state from which the move occurs), and also freezes a copy of the current counter of $\mathcal{A}$ in `c_freeze_1`. Then, upon another move between the same two segments, $\mathcal{V}$ may check whether the current state is again $q$. If so, $\mathcal{V}$ records the counter at the second move as well in `c_freeze_2`, at which point the simulation moves to the *2-NFA Phase*.

The *2-NFA Phase* assumes (and later verifies) that during its run, $\mathcal{A}$ encountered a positive loop, i.e., a path from configuration $(q, c, p)$ to $(q, c', p)$ with $c' > c$. In this case, $\mathcal{A}$ can charge the counter to an arbitrarily large value. Then, the word is accepted if it is accepted in the 2-NFA obtained from $\mathcal{A}$ by ignoring the counter, starting from $(q, p)$. In this phase we repeat the simulation phase above, ignoring the counter of $\mathcal{A}$, but still limiting to $|Q|$ moves between each two segments, with the reasoning that more than $|Q|$ moves imply a cycle, and therefore if there is an accepting run, there is also a shorter one.

Finally, upon encountering the accepting state of $\mathcal{A}$, we move to the *finalizing component*. There, it remains to verify that if the 2-NFA phase was reached, then indeed a positive loop was encountered. To do so, the two frozen counter values of the loop (`c_freeze_1`, `c_freeze_2`) are decreased simultaneously, and at some point a nondeterministic transition decreases only the (presumably higher) counter `c_freeze_2` and moves to the state $s_{\text{fin}}$. Then, all the counters are allowed to be decreased, so that we can reach $(s_{\text{fin}}, \mathbf{0})$. This ensures that $(s_{\text{fin}}, \mathbf{0})$ only if `c_freeze_2` was indeed higher than `c_freeze_1`, witnessing that a positive loop was encountered.

## 3.3 Emptiness of Sweeping 2-OCNs is Decidable

In this section we consider the sweeping fragment of 2-OCNs. A run $\rho$ of a 2-OCN is said to be *sweeping* if the head shift of the transitions changes only at the end-markers (i.e., the run "sweeps" forward from $\vdash$ to $\dashv$, and then sweeps backward to $\vdash$). A 2-OCN is *sweeping* if all its runs are sweeping. We assume w.l.o.g. that a sweeping 2-OCN only accepts when reading $\vdash$ (otherwise we add states to complete a backward sweep and then accept). We show that for sweeping 2-OCN, the emptiness problem is Ackermann-complete. The proof is split to the upper bound (Theorem 6) and lower bound (Theorem 7).

▶ **Theorem 6.** *The emptiness problem for sweeping 2-OCNs is decidable in Ackermannian complexity.*

**Proof.** We obtain the Ackermann upper bound of Theorem 6 by reducing nonemptiness of Sweeping 2-OCN to VASS-reachability. Similarly to Theorem 5, our proof relies on the observation that once a positive loop is encountered, we can replace the 2-OCN with a Boolean automaton. The challenge here, however, is that the language is not bounded, and therefore we cannot keep track of the head in a fixed number of segments. Instead, we use the sweeping property to simulate all the sweeps of the 2-OCN within a single pass, in a radically different approach than in the previous section.

Specifically, Consider a sweeping 2-OCN $\mathcal{A} = \langle Q, \Sigma, \Delta, q_{\mathsf{init}}, q_{\mathsf{acc}} \rangle$. We obtain from $\mathcal{A}$ a VASS with bounded 0-tests $\mathcal{V}$ such that $L(\mathcal{A}) \neq \emptyset$ iff $(s_{\mathsf{init}}, \mathbf{0})$ reaches $(s_{\mathsf{fin}}, \mathbf{0})$ for certain states $s_{\mathsf{init}}, s_{\mathsf{fin}}$ of $\mathcal{V}$. Moreover, this can be done in single-exponential time, and the size of $\mathcal{V}$ is single-exponential in that of $\mathcal{A}$.

The formal construction appears in Appendix B. We present the intuition here.

## Construction of $\mathcal{V}$ – Intuitive Overview

As an example, assume that $\mathcal{A}$ accepts a word $w \in \Sigma^*$ using two forward sweeps and two backward sweeps, as depicted in Fig 3.

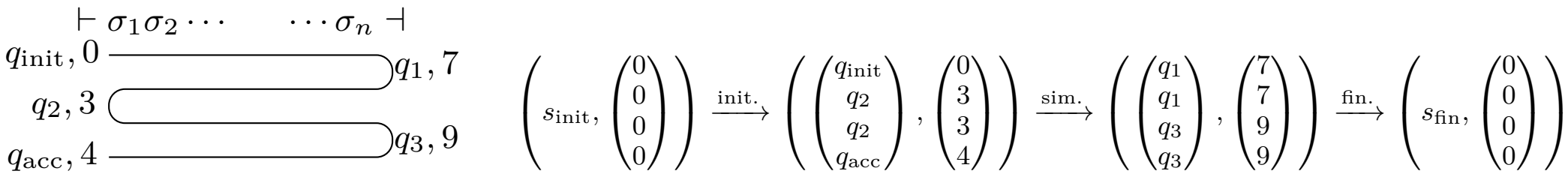

**Figure 3** Left: A run of a sweeping 2-OCN. The pairs represent the state and counter value at the end of each sweep. Right: The corresponding successful run of $\mathcal{V}$.

We can simulate the entire run of $\mathcal{A}$ on $w$ using a VASS $\mathcal{V}$ by, intuitively, tracking all the forward sweeps in parallel with all the backward sweeps, where for the backward sweeps we simulate the *reverse* 2-OCN of $\mathcal{A}$, denoted $\mathcal{A}^R$. The latter is a 2-OCN obtained from reversing the transitions, as well as negating the counter effects and the head shifts in all transitions. In order to make sure the sweeps concatenate correctly, we guess the states and the counter values. Specifically, we proceed as follows. We start by guessing the state in which each backward sweep ends. In this case, a successful guess is $(q_{\mathsf{init}}, q_2, q_{\mathsf{acc}})$ (note that all guesses must start and end in the initial and accepting state). We then guess the counter values with which each backward sweep ends. We store two copies of each guess, one used to simulate the backward run, and the other to simulate the next forward run (except for the last sweep). In our example, the successful guess is $(0, 3, 3, 4)$ (where 0 is the initial counter for the first forward sweep).

At each step, we now nondeterministically guess letters and corresponding transitions from $\mathcal{A}$ for the forward sweep, and from $\mathcal{A}^R$ for the backward sweep. After guessing the correct word $w$ and keeping track of the counters in all components, we would then reach the counters $(7, 7, 9, 9)$. Indeed, the forward run from 0 reaches counter 7, whereas the backward run from 3 starts from counter 7, and hence $\mathcal{A}^R$ would yield 7 along the reversal of the run. In addition, we track the states of $\mathcal{A}$ and of $\mathcal{A}^R$ along each run.

The last component of $\mathcal{V}$ verifies that the concatenation is correct, i.e., that each pair of adjacent counters are equal, and that the reached states match. The latter is encoded in the states, and the former is tested by decreasing both counters together, and checking reachability of $\mathbf{0}$.

In order to generalize the example above, it is crucial that the number of sweeps be *uniformly bounded*. To this end, we observe that if there are more than $|Q|$ forward-backward sweeps along a run, then a state is visited twice at $\vdash$, i.e., forms a loop. If this loop decreases the counter, then there is a shorter run with higher counters, so we need not simulate this former run. If this loop increases the counter, then by repeating it we can obtain an arbitrarily high counter. We can then discard the counter and look for a short accepting run, based on the 2-NFA obtained from $\mathcal{A}$, similarly to Section 3.2. However, unlike Section 3.2, we cannot simply simulate the 2-NFA, since we are committed to a single sweep on the guessed word. In order to overcome this, we *initially* guess which state $q$ will repeat in a positive loop, and then simulate a DFA that is equivalent [31, 30, 32] to the 2-NFA above, starting from state $q$. At the end of the run, we verify using the reachability query that the loop is indeed positive. Combining these ideas yields a VASS with the conditions above. ◀

▶ **Theorem 7.** *The emptiness problem for sweeping 2-OCN is Ackermann-hard.*

**Proof.** We show a reduction from VASS reachability, known to be Ackermann-hard [23, 9], to nonemptiness of sweeping 2-OCN. Specifically, we assume that the reachability query is from $(s_1, \mathbf{0})$ to $(s_2, \mathbf{0})$, as this preserves the hardness.

Consider a VASS $\mathcal{V} = \langle S, T \rangle$ of dimension $d$ and configurations $(s_1, \mathbf{0})$ to $(s_2, \mathbf{0})$. We construct a sweeping 2-OCN $\mathcal{A} = \langle Q, T, \Delta, q_{\text{init}}, q_{\text{acc}} \rangle$ where the alphabet is $T$, i.e., the transitions of $\mathcal{V}$. $\mathcal{A}$ works as follows: given a word $w \in T^*$, first $\mathcal{A}$ sweeps forward and checks that the states given by the transitions follow a run from $s_1$ to $s_2$ in $\mathcal{V}$ (ignoring the counter values). If this fails, $\mathcal{A}$ rejects.

Next, $\mathcal{A}$ checks that the counter updates are valid and take the configuration from $\mathbf{0}$ to $\mathbf{0}$. This is done in two phases. In the first phase, $\mathcal{A}$ makes $d$ forward sweeps, where sweep $i$ simulates the $i$-th counter of $\mathcal{V}$ in the transitions given by $w$, and each backward sweep resets the counter to 0 by negating the effect of the transition. Thus, if the $d$ sweeps complete successfully, we are assured that $w$ prescribes a run that reaches from $(s_1, \mathbf{0})$ to $(s_2, \boldsymbol{v})$ for some $\boldsymbol{v} \geq \mathbf{0}$. Then, $\mathcal{A}$ makes a forward sweep while keeping the counter at 0 and starts the second phase, where we perform the same simulation but backwards: we check that the run prescribed by $w^R$ (the reverse of $w$) leads from $(s_2, \mathbf{0})$ to $(s_1, \boldsymbol{v'})$ for some $\boldsymbol{v'} \geq \mathbf{0}$ (note that here we negate the counter operations).

The main observation required to complete the proof is that a for a run $\rho$ of $\mathcal{V}$ from $(s_1, \mathbf{0})$ to $(s_2, \boldsymbol{v})$ we have that $\rho^R$ (with negated counters) is a valid run from $(s_2, \mathbf{0})$ iff $\boldsymbol{v} = 0$. This follows immediately from the VASS semantics.

We conclude that $L(\mathcal{A}) \neq \emptyset$ iff $(s_2, \mathbf{0})$ is reachable from $(s_1, \mathbf{0})$ in $\mathcal{V}$. ◀

## 3.4 The Membership Problem for 2-OCN is Polytime

Recall that by [7], if a 2-OCN $\mathcal{A}$ has counter updates in $\{-1, 0, 1\}$, we can compute in polynomial time (in the description of $\mathcal{A}$) a polynomial $p$ such that if a word $w$ is accepted, then it is accepted with a run of length at most $p(|w|)$. Unfortunately, this does not carry through to our setting, where counters are encoded in binary, as the following example shows:

Nevertheless, we show that 2-OCN retain efficient membership even with binary counters:

▶ **Theorem 8.** *The membership problem for 2-OCN (given a 2-OCN $\mathcal{A}$ and a word $w$, is $w \in L(\mathcal{A})$?) is decidable in polynomial time.*

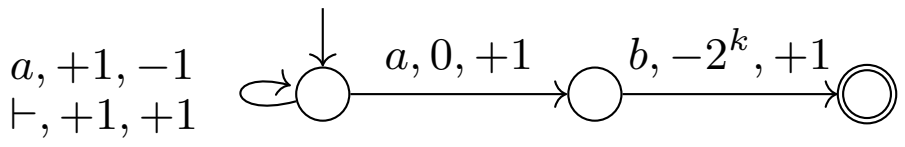

**Figure 4** A 2-OCN $\mathcal{A}_k$ parameterized by $k \in \mathbb{N}$. The word $\vdash ab \dashv$ is accepted, but the shortest accepting run is of length $2^k + 1$, since the counter "charges" for $2^k$ steps reading only the $\vdash a$ prefix.

**Proof.** Consider a 2-OCN $\mathcal{A} = \langle Q, \Sigma, \Delta, q_{\text{init}}, q_{\text{acc}} \rangle$ and a word $w \in \Sigma^*$. We implicitly assume that $w$ already includes the end-markers. Similarly to the reasoning in Sections 3.2 and 3.3, we observe that if an accepting run of $\mathcal{A}$ on $w$ visits the same state/position pair $(q, p)$ twice, then if the counter effect between the visits is nonpositive, a shorter accepting run exists, and if the counter effect is positive, then this cycle can be pumped arbitrarily, so that acceptance is dependent only on the 2-NFA obtained by ignoring the counter effects starting from $(q, p)$. Since the number of state/position pairs is $|w| \cdot |Q|$, it follows that within $|w| \cdot |Q| + 1$ steps we visit a cycle. Similarly, in the 2-NFA if an accepting run exists, then there is also one of length at most $|w| \cdot |Q|$.

Deciding whether there exists an accepting run in the 2-NFA from $(q, p)$ is straightforward: we keep track of the reachable set of configurations $(q', p')$ at each step of the run, up to $|w| \cdot |Q|$ steps, and if $q_{\text{acc}}$ is reached then the run is accepting.

It remains to detect acceptance in the 2-OCN, or to detect a positive cycle. To do so, we proceed as follows. We simulate $\mathcal{A}$ for $|w| \cdot |Q|$ steps, and for each step $1 \leq i \leq |w| \cdot |Q|$ we store a function $f_i \colon Q \times \{1, \ldots, |w|\} \to \mathbb{N}$ that stores for each configuration $(q, p)$ the maximal counter with which $(q, p)$ can be reached after $i$ steps. It is easy to compute $f_i$ from $f_{i-1}$ in polynomial time. Once this simulation is complete, if $q_{\text{acc}}$ does not appear, then we want to look for a positive cycle. To this end, for each $1 \leq j \leq |w| \cdot |Q|$ and configuration $(q, p)$, we again construct functions $g_j$ similarly to the above, with initial configuration $f_i(q, p)$. If at any point we encounter $(q, p)$ with a counter higher than $f_i(q, p)$, then we have a positive cycle. Otherwise, the word is not accepted.

It is easy to verify that all the computations can be carried out in polynomial time. ◀

## 4 On the Semilinearity of 2-OCN Languages

**Preliminaries.** Emptiness decidability has been historically intimately tied to *semilinearity*. A set $E \subseteq \mathbb{N}^d$ is semilinear if it is expressible in first-order logic with addition (there are many equivalent definitions and we will not need a specific one). The *Parikh image* of a language $L$ over an alphabet $\{a_1, \ldots, a_k\}$ is the set of vectors $(n_1, \ldots, n_k)$ such that there is a word in $L$ with precisely $n_i$ letters $a_i$ (in any order), for all $i$. In other words, the Parikh image counts the number $|w|_\ell$ of times each letter $\ell$ appears in a word $w \in L$, resulting in a set of vectors in $\mathbb{N}^{|\Sigma|}$. A class of languages is *semilinear* if for any language $L$ in that class, the Parikh image of $L$ is semilinear. In that case, if there is an algorithmic way to obtain a representation of that semilinear set from a representation of $L$, then one can check whether $L = \emptyset$, since satisfiability is decidable for first-order logic with addition [6].

**General 2-OCN.** Since the emptiness problem is undecidable for 2-DOCN, it is an unsurprising fact that there are nonsemilinear 2-DOCN languages (e.g., $\{(a^n\#)^n \mid n > 1\}$). A tantalizing question surging from the study in Section 3.2 is: *Are all 2-OCN bounded languages semilinear?* For 2-DOCN, it is proved in [26] that this is indeed the case. For 2-OCN, it is likely that a proof of this statement would provide an alternative proof to Theorem 5. We conjecture that, indeed, 2-OCN are in good company with a wealth of computational models whose bounded languages are precisely those with a semilinear Parikh image. We come short of showing this, but prove, using the construction of Section 3.2 and a result of [15], that:

▶ **Theorem 9.** *For any 2-OCN $\mathcal{A}$ with $L(\mathcal{A}) \subseteq a_1^* a_2^* \cdots a_k^*$ and for any $\Gamma \subseteq \{a_1, a_2, \ldots, a_k\}$, the set $\left\{\sum_{\ell \in \Gamma} |w|_\ell \mid w \in L(\mathcal{A})\right\} \subseteq \mathbb{N}$ is effectively semilinear.*

**Proof.** In [15, Corollary 4.3] the following is shown.[2] Given a VASS $\mathcal{V} = \langle S, T\rangle$ of dimension $d$ and two states $s_1, s_2$, consider the set $\mathcal{R}_{s_1,s_2} \subseteq \mathbb{N}^d \times \mathbb{N}^d \times \mathbb{N}^T$ such that $(\boldsymbol{p}, \boldsymbol{p'}, \boldsymbol{t}) \in \mathcal{R}_{s_1,s_2}$ iff $(s_2, \boldsymbol{p'})$ is reachable from $(s_1, \boldsymbol{p})$ via a run $\rho$ that takes transition $\tau$ exactly $\boldsymbol{t}(\tau)$ times. Then the image of every morphism $\pi\colon \mathcal{R}_{s_1,s_2} \to \mathbb{N}$ is effectively semilinear.

In our setting, take $\mathcal{V}$ to be the VASS obtained in Section 3.2 with the states $s_{\text{init}}, s_{\text{fin}}$ and $\boldsymbol{p} = \boldsymbol{p'} = \boldsymbol{0}$. Let $\pi$ be the morphism that counts the number of transitions used to initialize $n_i$ for all $a_i \in \Gamma$, and recall that by the correctness of the construction of $\mathcal{V}$, we have that along a run from $(s_{\text{init}}, \boldsymbol{0})$ to $(s_{\text{fin}}, \boldsymbol{0})$, the value of $n_i$ after the Initialization Component corresponds to the length of the $a_i$ segment in an accepted word. Then, [15, Corollary 4.3] readily shows that the set $\left\{\sum_{a_i \in \Gamma} n_i \mid a_1^{n_1} \cdots a_k^{n_k} \in L(\mathcal{A})\right\}$ is effectively semilinear. ◀

The previous theorem implies that 2-OCN languages over a singleton alphabet are semilinear. In addition, we have the following:

▶ **Corollary 10.** *The languages $\{0^n 1^{n^2} \mid n \geq 1\}$ and $\{0^n 1^{2^n} \mid n \geq 1\}$, known to be 2-DOCA languages [29], are not 2-OCN languages.*

**Sweeping 2-OCN.** We note that Theorem 9 also applies to sweeping 2-OCN without the bounded language restriction, with a similar proof.

▶ **Theorem 11.** *For any sweeping 2-OCN $\mathcal{A}$ over $\Sigma$ and $\Gamma \subseteq \Sigma$, $\left\{\sum_{\ell \in \Gamma} |w|_\ell \mid w \in L(\mathcal{A})\right\}$ is effectively semilinear.*

However, unlike the case of bounded languages, here we can show that the Parikh image of sweeping 2-OCN language are not always semilinear.

▶ **Proposition 12.** *There are nonsemilinear languages recognized by sweeping 2-DOCN.*

**Proof.** Consider the Dyck language over $\{a, b\}$, i.e., well-parenthesized expressions where $a$ corresponds to an opening parenthesis and $b$ to a closing one. Write $P$ for the prefixes of that language. Clearly, $P$ is a 1-DOCN language. Write $P'$ for the language $P$ where $a$ and $b$ swap roles. We can construct a sweeping 2-DOCN that recognizes $K = P \cap a^* P'$ using the closure of 2-DOCN under intersection [26] (which preserves the sweeping property). In any word $w$ in $K$, if $n$ is the length of the first block of $a$'s, then $0 \leq |w|_a - |w|_b \leq n$. Finally, using a few more sweeps, we can express:

$$L = \{c^n \# w \mid w \in K \wedge w \in a^n (b^+ a^+)^{n-1} b^n\}.$$

Indeed, this requires checking that the number of $c$'s is equal to the number of $a$'s in the first block, the number of $b$'s in the last block, and the number of $ab$ infixes.

Now, for a fixed value $n$, the longest word in $L$ that can appear after $c^n\#$ is $(a^n b^n)^n$, of length $2n^2$, showing that this language is not semilinear. ◀

[2] In [15] the result is phrased for Petri Nets, but the equivalence between Petri Nets and VASS gives a straightforward translation. In [22], a slightly weaker result is presented in the language of VAS.

We will show, in Section 5, that there is a 1-OCN language (thus a sweeping 2-OCN language) this is not expressible using a 2-DOCN. Conversely, Proposition 12 provides an easy way to show that:

▶ **Corollary 13.** *There is a 2-DOCN language that is not expressible using a sweeping 2-OCN.*

**Proof.** Consider the language $L$ over the alphabet $\{a, b, c, \#, \underline{b}, x\}$ defined as:

$$L = \{a^n \# b^m x^m \# \underline{b}^m x^{2m} \# \underline{b}^m x^{3m} \# \cdots \# \underline{b}^m c^{nm} \mid n, m > 0\}$$

By applying the morphism $h \colon \Sigma^* \to \Sigma^*$ that erases $\underline{b}, \#$, and $x$, we have $h(L) = \{a^n b^m c^{nm}\}$, which can be seen as multiplication.

The language $L$ is recognizable by a 2-DOCN $\mathcal{A}$ by repeatedly using a similar approach to Example 1:

- Check that the input has the form $a^* \# b^* (x^* \# \underline{b}^*)^* c^*$,
- Check that the number of $a$'s is the same as the number of $\#$'s,
- Check that the first sequence of $\underline{b}$'s is as long as the sequence of $b$'s,
- Check that the lengths of all sequences of $\underline{b}$'s are equal, by checking that each neighboring sequence $\#\underline{b}^i x^* \# \underline{b}^j x^*$ satisfies $i = j$,
- Check for each sequence $x^i \# \underline{b}^m x^j$ that $i + m = j$, and the same for the last segment where $x$ is replaced with $c$.

This 2-DOCN proceeds segment by segment, always making sure that the number of $x$'s increases by exactly $m$. This, together with the initial test that the number of segments is $n$, forces the number of $c$'s to be exactly $nm$.

However, the lengths of words in this language do not form a semilinear set, so by Theorem 11 it cannot be recognized by a sweeping 2-OCN. ◀

Over bounded languages, the following exact characterization holds:

▶ **Theorem 14.** *A bounded language is recognized by a sweeping 2-OCN iff it is semilinear.*

**Proof.** The fact that any semilinear bounded language can be expressed with a sweeping 2-OCN is easily deduced from the fact that semilinear sets are those expressible with *quantifier-free* first-order formulas with addition, order, and modulo. Putting such a formula in DNF, we simply need to guess which conjunction is to hold, and the conjunction can be checked by a sweeping 2-DOCN using the closure of 2-DOCN under intersection [26] (which preserves the sweeping property).

We turn to the fact that bounded languages expressed by sweeping 2-OCN are semilinear. We split the proof into two parts.

**Constant number of sweeps.** Let us first assume that the sweeping 2-OCN $\mathcal{A}$ executes a constant number of sweeps, say $C$, with $C$ odd, for each input word (we assume here that the machine ends at the right end-marker). With $L(\mathcal{A}) \subseteq a_1^* \cdots a_k^*$, consider the following transformation that produces a word over the alphabet $\Gamma = \{a_{i,j} \mid 1 \leq i \leq C \wedge 1 \leq j \leq k\}$:

$$w = a_1^{n_1} \cdots a_k^{n_k} \quad \rightsquigarrow \quad w' = a_{1,1}^{n_1} \cdots a_{1,k}^{n_k} \# a_{2,k}^{n_k} \cdots a_{2,1}^{n_1} \# \cdots \# a_{C,1}^{n_1} \cdots a_{C,k}^{n_k}.$$

That is, the input is replicated $C$ times, with every other copy reversed, and each copy on its own alphabet. Let us call $w'$ the *replica* of $w$. Since $\mathcal{A}$ makes at most $C$ sweeps, we can build a 1-OCN $\mathcal{B}$ that reads the replica of any word $w \in a_1^* \cdots a_k^*$ and accepts if and only if $w$ is accepted by the 2-OCN (and the input is of the format $R = a_{1,1}^* \cdots a_{1,k}^* \# a_{2,k}^* \cdots a_{2,1}^* \# \cdots \# a_{C,1}^* \cdots a_{C,k}^*$). Naturally, $\mathcal{B}$ also accepts words that are *not* replicas.

We circumvent this problem by using an "external" semilinear set, as follows. Let $E$ be the Parikh image of the language where for all $i$, each $a_{\bullet,i}$ appears the same number of times; clearly, a word in $R$ is a replica iff its Parikh image is in $E$. Since $\mathcal{B}$ is a 1-OCN language (and in particular context-free), the Parikh image $F$ of its language is semilinear by Parikh's Theorem [28]. It is then easy to check that $E \cap F$ is exactly the set of Parikh images of replicas of words in $L(\mathcal{A})$, showing that the Parikh image of $L(\mathcal{A})$ itself is semilinear (as a projection of the former).

**Any number of sweeps.** If $\mathcal{A}$ is sweeping but does not do a constant number of sweeps for each input, we adapt the recurring idea of Sections 3.2 and 3.3. Let $k$ be the number of states of $\mathcal{A}$. After $2k + 1$ sweeps, a state is repeated at the left end-marker; if the counter effect between the two repetitions is nonpositive, then that run is not useful for acceptance, and can be disregarded. If the counter *increases*, then it can be increased arbitrarily, and the behavior of the 2-OCN is the same as a 2-NFA $\mathcal{N}$ from that point on. Consider the 1-OCN $\mathcal{B}$ constructed above, with $C = 2k + 1$. We can additionally have $\mathcal{B}$ guess the state that repeats, if any, and check that it indeed repeats. Next, $\mathcal{B}$ assumes that the counter strictly increases in the repetition, and branches into a DFA for $\mathcal{N}$. In order to "kill" the runs of $\mathcal{B}$ where this assumption is incorrect, any transition that $\mathcal{B}$ takes, in between the first and second appearance of the repeated state, should be followed by input symbols indicating the counter effect, increasing ($\texttt{+}^*$) or decreasing ($\texttt{-}^*$), and by how much. $\mathcal{B}$ then verifies that the correct effect is matched to the transition of $\mathcal{A}$ it guesses. For instance, $\mathcal{B}$ may read the following replica of $aabbbc$ with extra counter information (here the original alphabet is $\{a, b, c\}$ and the replica alphabet has $1, \ldots, 5$ as indices):

$$\underbrace{a_1a_1b_1b_1b_1c_1\#c_2b_2b_2b_2a_2a_2}_{\mathcal{B}\text{ simulates two sweeps of }\mathcal{A}}\ \underset{\uparrow}{\#}\ \underbrace{a_3\texttt{++}a_3\texttt{+}b_3\texttt{-}b_3\texttt{+}b_3\texttt{-}c_3\texttt{+}\#c_4\texttt{-}b_4\texttt{+}b_4\texttt{+}b_4\texttt{-}a_4\texttt{-}a_4\texttt{-}}_{\text{counter updates appear in input showing action of }\mathcal{A}\text{'s transitions}}\ \underset{\uparrow}{\#}\ \underbrace{a_5a_5b_5b_5b_5c_5\#}_{\text{A DFA for }\mathcal{N}\text{ is ran}}$$

$\mathcal{B}$ guesses state repetition

$\mathcal{B}$ checks state repetition
and assumes that counter increased

As mentioned, this modified version of $\mathcal{B}$ may jump into a DFA for $\mathcal{N}$ when it should not. However, we can augment the external semilienar set $E$, checking that the input word is a replica, to also check that the number of $\texttt{+}$ is strictly greater than the number of $\texttt{-}$. This corresponds to the counter updates between the state repetitions having a strict positive impact, and that $\mathcal{B}$ correctly jumps into $\mathcal{N}$'s simulation. To conclude, let again $F$ be the (semilinear) Parikh image of the language of $\mathcal{B}$. The vectors in $E \cap F$ correspond precisely to the replicas (with possibly explicit counter updates) that are accepted by $\mathcal{A}$, hence the language of $\mathcal{A}$ is semilinear. ◀

## 5 2-DOCN vs 1-OCN: A Pumping Lemma for 2-DOCN

In [7] it is shown that the language $\{a^n b^m \mid n \neq m\}$ is 2-OCN recognizable but not 2-DOCN recognizable. We strengthen this result in two directions: first we present a general pumping lemma that we can then use to establish that various languages are not 2-DOCN-recognizable. Second, using our lemma we are able to find a language that is not 2-DOCN recognizable, but is already 1-OCN recognizable. This shows that nondeterminism can be used to compensate for two-wayness, in some settings.

▶ **Lemma 15.** *Let $\Sigma$ be an alphabet and $a \in \Sigma$. For every 2-DOCN-recognizable language $L \subseteq \Sigma^*$ there exists $K \in \mathbb{N}$ such that for every $x, y \in \Sigma^*$, if $xa^K y \notin L$ then there exists $K' \neq K$ such that $xa^{K'} y \notin L$ as well.*

We give a rough sketch of the main ideas of the proof, see Appendix C for the complete arguments and definitions. Consider a 2-DOCN $\mathcal{A} = \langle Q, \Sigma, \Delta, q_{\text{init}}, q_{\text{acc}} \rangle$, and think of $K$ as some large constant. Our goal is to devise a way to pump certain infixes of a word $xa^K y$, so that we can reason about the resulting run of $\mathcal{A}$ on it. Naturally, the challenge lies in synchronizing the behaviors of forward and backward head movements in the run.

We think of the $a^K$ infix as partitioned to three parts: two short $a^{|Q|}$ segments referred to as *outer left a's* and *outer right a's*, and a long $a^{K-2|Q|}$ segment in the middle, called the *inner a's*. We depict the word $xa^K y$ as 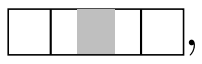, with the grayed area representing the inner $a$'s, and the vertical lines marking the end of $x$ and the beginning of $y$.

We then consider the form of the run of $\mathcal{A}$ on $xa^K y$. Specifically, we divide runs according to whether they cross into the inner $a$'s. We show that runs that do reach the inner $a$'s, must continue to the end of the tape (either from right or left). Runs that do not cross the inner $a$'s must therefore get stuck in a small part of the tape. By depicting runs as going from top to bottom upon head reversal, we have e.g., the type 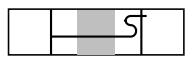 which depicts runs going left, and 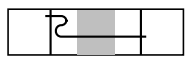 depicts runs going right. The possible types of runs are then *left loop* (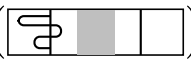), *left sink* (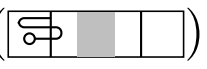), *left crossing* (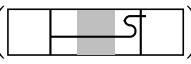) and their right counterparts. Each type signifies a specific behavior of the runs, e.g., whether they repeat a state, or stop due to the counter becoming negative.

We define a *round trip* to be a concatenation of runs of the form 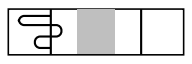 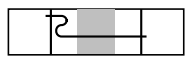 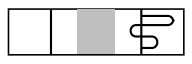 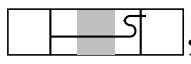, and show that either (1) $\mathcal{A}$ has a run on $xa^K y$ with many round-trips, or (2) it has a run with a few round trips that gets stuck in a loop, or (3) its maximal run has few round trips and ends with the counter becoming negative. We then provide pumping arguments according to each type of run.

A crux of the proof lies in handling Form (3) above. The main idea is the following: consider a run with several round-trips that ends due to a counter becoming negative. Since the inner $a$'s are a long infix, we can find two indices that are visited with the same sequence of states in the round trips. We can then attempt to "pump" or "cut" the infix between these indices, but this may cause the counter to ultimately increase, thus yielding an accepting run from a previously-rejecting one. We therefore carefully analyze the effect of the counter between these two indices on each pass of the form 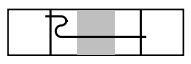 or 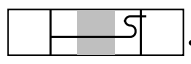. If at any point the cumulative sum of effects is negative, we can pump the run enough so that the counter ultimately becomes negative, and the word is not accepted. Otherwise, all the cumulative sums are non-negative, in which case we can cut the cycle, causing the counter to either become negative or to complete the run in the same state as the original run, which is not accepting. The precise details are involved and appear in Appendix C.

▶ **Corollary 16.** $L = \{a^\ell b^m c^n \mid \ell > m \vee m > n\}$ *is 1-OCN but not 2-DOCN recognizable.*

**Proof.** $L$ is 1-OCN recognizable by guessing whether $\ell > m$ or $m > n$, and checking by either increasing the counter on $a$ and decreasing on $b$, or increasing on $b$ and decreasing on $c$ (with another decrease at the end to make the inequality strict). Assume by way of contradiction that $L = L(\mathcal{D}_2)$ for a 2-DOCN $\mathcal{D}_2$. Let $K$ be the constant provided by Lemma 15, then $a^K b^K c^K \notin L = L(\mathcal{D}_2)$. Then, Lemma 15 guarantees that there exists $K' \neq K$ such that $a^K b^{K'} c^K \notin L(\mathcal{D}_2)$, but $a^K b^{K'} c^K \in L$, since either $K > K'$ or $K' > K$, a contradiction. ◀

We remark that we can similarly show that $L_2 = \{a^n b^m \mid n \neq m\}$ is also not 2-DOCN recognizable (but is recognizable by a sweeping 2-OCN), giving an alternative proof to the result in [26].

## 6 Research Directions

**Separations.** The classes we consider, 1-OCN, 1-DOCN, 2-OCN, 2-DOCN, sweeping and nonsweeping, over bounded languages or not, are all provably distinct, except for the following: are all bounded 2-OCN languages expressible with a sweeping machine? We conjecture that they are, and leave this question open (this is known for 2-DOCN from [7]).
**Closure.** The examples of languages outside 2-DOCN readily show that the class of languages recognized by 2-DOCN is not closed under union nor complement, but is closed under intersection (also shown in [26]). For 2-OCN, the class is closed under union, but we conjecture that it is closed under neither intersection nor complement. We leave these questions open, together with the same questions for sweeping 2-OCN.
**Decidability.** Beyond emptiness, our proofs imply that it is decidable whether the intersection of two (sweeping or bounded) 2-OCN is empty. The next natural question is then the decidability of *inclusion*. We conjecture that this problem is decidable for sweeping and bounded 2-OCN.

*Acknowledgments.* We are grateful to Dmitry Chistikov for shedding light on some claims made in [4] and to an anonymous reviewer for providing some key references.

### References

1 Shaull Almagor, Guy Avni, Henry Sinclair-Banks, and Asaf Yeshurun. Dimension-minimality and primality of counter nets. In *International Conference on Foundations of Software Science and Computation Structures*, volume 14575, pages 229–249. Springer, Springer, 2024. URL: `https://doi.org/10.1007/978-3-031-57231-9_11`, `doi:10.1007/978-3-031-57231-9_11`.
2 Shaull Almagor, Udi Boker, Piotr Hofman, and Patrick Totzke. Parametrized universality problems for one-counter nets. In *31st International Conference on Concurrency Theory (CONCUR 2020)*. Schloss Dagstuhl-Leibniz-Zentrum für Informatik, 2020. `doi:10.4230/LIPIcs.CONCUR.2020.47`.
3 Shaull Almagor and Asaf Yeshurun. Determinization of one-counter nets. In *33rd International Conference on Concurrency Theory (CONCUR 2022)*. Schloss Dagstuhl-Leibniz-Zentrum für Informatik, 2022. `doi:10.4230/LIPIcs.CONCUR.2022.18`.
4 Setsuo Arikawa. On some properties of length-growing functions on two-way pushdown automata. *Memoirs of the Faculty of Science, Kyushu University. Series A, Mathematics*, 22(2):110–127, 1968.
5 Mohamed Faouzi Atig and Pierre Ganty. Approximating petri net reachability along context-free traces. In *IARCS Annual Conference on Foundations of Software Technology and Theoretical Computer Science (FSTTCS)*, pages 152–163. Dagstuhl, Germany: Leibniz-Zentrum für Informatik, 2011. `doi:10.4230/LIPIcs.FSTTCS.2011.152`.
6 Leonard Berman. The complexity of logical theories. *Theoretical Computer Science*, 11(1):71–77, 1980. `doi:10.1016/0304-3975(80)90037-7`.
7 Tat-hung Chan. On two-way weak counter machines. *Mathematical systems theory*, 20(1):31–41, 1987. `doi:10.1007/BF01692057`.
8 Wojciech Czerwiński, Sławomir Lasota, Ranko Lazić, Jérôme Leroux, and Filip Mazowiecki. The reachability problem for petri nets is not elementary. *Journal of the ACM (JACM)*, 68(1):1–28, 2020. `doi:10.1145/3422822`.
9 Wojciech Czerwiński and Łukasz Orlikowski. Reachability in vector addition systems is ackermann-complete. In *2021 IEEE 62nd Annual Symposium on Foundations of Computer Science (FOCS)*, pages 1229–1240. IEEE, 2022. `doi:10.1109/FOCS52979.2021.00120`.
10 Zhe Dang, Oscar H Ibarra, and Zhi-Wei Sun. On two-way nondeterministic finite automata with one reversal-bounded counter. *Theoretical computer science*, 330(1):59–79, 2005. `doi:10.1016/j.tcs.2004.09.010`.

11 Marzio De Biasi and Abuzer Yakaryılmaz. Unary languages recognized by two-way one-counter automata. In *International Conference on Implementation and Application of Automata*, pages 148–161. Springer, 2014. doi:10.1007/978-3-319-08846-4_11.

12 Pavol Duris and Zvi Galil. Fooling a two way automation or one pushdown store is better than one counter for two way machines. *Theoretical Computer Science*, 21(1):39–53, 1982. doi:10.1016/0304-3975(82)90087-1.

13 Zvi Galil. Some open problems in the theory of computation as questions about two-way deterministic pushdown automaton languages. *Mathematical systems theory*, 10(1):211–228, 1976. doi:10.1007/BF01683273.

14 James N Gray, Michael A Harrison, and Oscar H Ibarra. Two-way pushdown automata. *Information and Control*, 11(1-2):30–70, 1967. doi:10.1016/S0019-9958(67)90369-5.

15 Dirk Hauschildt and Matthias Jantzen. Petri net algorithms in the theory of matrix grammars. *Acta Informatica*, 31:719–728, 1994. doi:10.1007/BF01178731.

16 Piotr Hofman, Richard Mayr, and Patrick Totzke. Decidability of weak simulation on one-counter nets. In *2013 28th Annual ACM/IEEE Symposium on Logic in Computer Science*, pages 203–212. IEEE, 2013. doi:10.1109/LICS.2013.26.

17 Piotr Hofman and Patrick Totzke. Trace inclusion for one-counter nets revisited. In *Reachability Problems: 8th International Workshop, RP 2014, Oxford, UK, September 22-24, 2014. Proceedings 8*, pages 151–162. Springer, 2014. doi:10.1007/978-3-319-11439-2_12.

18 Oscar H Ibarra and Zhe Dang. On two-way fa with monotonic counters and quadratic diophantine equations. *Theoretical computer science*, 312(2-3):359–378, 2004. doi:10.1016/j.tcs.2003.10.027.

19 Oscar H Ibarra, Tao Jiang, Nicholas Tran, and Hui Wang. On the equivalence of two-way pushdown automata and counter machines over bounded languages. In *STACS 93: 10th Annual Symposium on Theoretical Ascpects of Computer Science Würzburg, Germany, February 25–27, 1993 Proceedings 10*, pages 354–364. Springer, 1993. doi:10.1007/3-540-56503-5_36.

20 Oscar H Ibarra, Tao Jiang, Nicholas Tran, and Hui Wang. New decidability results concerning two-way counter machines. *SIAM Journal on Computing*, 24(1):123–137, 1995. doi:10.1137/S0097539792240625.

21 Oscar H Ibarra and Jianwen Su. Counter machines: decision problems and applications. In *Jewels are Forever: Contributions on Theoretical Computer Science in Honor of Arto Salomaa*, pages 84–96. Springer, 1999.

22 Hans Kleine Büning, Theodor Lettmann, and Ernst W. Mayr. Projections of vector addition system reachability sets are semilinear. *Theoretical Computer Science*, 64(3):343–350, 1989. URL: https://www.sciencedirect.com/science/article/pii/0304397589900558, doi:10.1016/0304-3975(89)90055-8.

23 Jérôme Leroux. The reachability problem for petri nets is not primitive recursive. In *2021 IEEE 62nd Annual Symposium on Foundations of Computer Science (FOCS)*, pages 1241–1252. IEEE, 2022. doi:10.1109/FOCS52979.2021.00121.

24 Daniel Martin and John Gwynn. Two results concerning the power of two-way deterministic pushdown automata. In *Proceedings of the ACM annual conference*, pages 342–344, 1973. doi:10.1145/800192.805729.

25 Marvin Lee Minsky. *Computation.* Prentice-Hall Englewood Cliffs, 1967.

26 Satoru Miyano. Two-way deterministic multi-weak-counter machines. *Theoretical Computer Science*, 21(1):27–37, 1982. doi:10.1016/0304-3975(82)90086-X.

27 Burkhard Monien. Deterministic two-way one-head pushdown automata are very powerful. *Information processing letters*, 18(5):239–242, 1984. doi:10.1016/0020-0190(84)90001-2.

28 Rohit J. Parikh. On context-free languages. *J. ACM*, 13(4):570–581, 1966. doi:10.1145/321356.321364.

29 Holger Petersen. Two-way one-counter automata accepting bounded languages. *ACM SIGACT News*, 25(3):102–105, 1994. doi:10.1145/193820.193835.

30 Michael O Rabin and Dana Scott. Finite automata and their decision problems. *IBM journal of research and development*, 3(2):114–125, 1959. doi:10.1147/rd.32.0114.

31 John C Shepherdson. The reduction of two-way automata to one-way automata. *IBM Journal of Research and Development*, 3(2):198–200, 1959. doi:10.1147/rd.32.0198.

32 Moshe Y. Vardi. A note on the reduction of two-way automata to one-way automata. *Information Processing Letters*, 30(5):261–264, 1989. doi:10.1016/0020-0190(89)90205-6.

# A Detailed Construction of $\mathcal{V}$ in Section 3.2

We follow the intuition given in Section 3.2. Let $\mathcal{A} = \langle Q, \Sigma, \Delta, q_{\text{init}}, q_{\text{acc}} \rangle$. Recall that $\Sigma = \{a_1, \ldots, a_k\}$. When constructing $\mathcal{V}$ we refer to vector entries as *counters*, and we specify vector operations on individual counters, e.g., if $c$ is a specific counter, then $c \leftarrow +1$ refers to the vector that is all 0 apart from +1 in the entry corresponding to $c$. Similarly, we can combine several counter operations in a single vector (e.g., $c_1 \leftarrow +1, c_2 \leftarrow -1$).

We start by describing the counters of $\mathcal{V}$ and their intuitive roles.

- `c` – tracking the counter of $\mathcal{A}$
- `c_freeze_1` – a copy of `c`, to be frozen at the beginning of a positive loop.
- `c_freeze_2` – a copy of `c`, to be frozen at the end of a positive loop.
- `L[1]`...`L[k]` and `R[1]`...`R[k]` – a pair of counters marking the position of the head from the Left and from the Right of each segment.

We now describe the states and transitions of $\mathcal{V}$ by listing the information stored in each state, and its intuitive meaning. We divide the states according to the components of $\mathcal{V}$.

**Initialization Component.** This component consists of a single state $s_{\text{init}}$. The transitions include a $k$ self loops for $i \in \{1, \ldots, k\}$, each with operation $R[i] \leftarrow +1$. These transitions allow $\mathcal{V}$ to "charge" the number of letters in each $a_i$ segment. Another transition (with operation $\mathbf{0}$) leads to the simulation component with current state $q_{\text{init}}$ and head position on $\vdash$ (i.e., current segment 0, see below).

**Simulation Component.** A state contains the following information:

- `cur_state` $\in Q$ – the current state of $\mathcal{A}$.
- `cur_seg` $\in \{0, \ldots, k+1\}$ – the current segment, with 0 and $k+1$ representing $\vdash$ and $\dashv$, respectively.
- `num_moves`: $\{(i, i+1), (i+1, i) \mid i \in \{0, \ldots, k\}\} \to \{0, \ldots, |Q|+1\}$ – how many times did we move from segment $i$ to segment $i+1$ (bounded by $|Q|+1$).
- `loop_seg_move` $\in \{(i, i+1), (i+1, i) \mid i \in \{0, \ldots, k\}\} \cup \{\bot\}$ – the border between segments where we guess a loop occurs, or $\bot$ if this is not guessed yet.
- `loop_state` $\in Q \cup \{\bot\}$ – the state where we guess a loop occurs, or $\bot$ if this is not guessed yet.

The transitions within segments are induced by those of $\mathcal{A}$: if `cur_state` $= q$ and `cur_seg` $= i$, then each transition $(q, a_i, e, h, q') \in \Delta$ induces a transition to a state with `cur_state` $= q'$ and the counter operations `c` $\leftarrow +e$, `c_freeze_2` $\leftarrow +e$, `L[i]` $\leftarrow +h$, `R[i]` $\leftarrow -h$. Additionally, if `loop_seg_move` $=$ `loop_state` $= \bot$ then `c_freeze_1` behaves like `c`.

The transition between segments are similarly induced, but instead of updating `L[i]` and `R[i]`, if $h = 1$ then $\mathcal{V}$ guesses that we move to a state with $cur_seg = i+1$. Then, `R[i]` is 0-tested and `num_moves`$(i, i+1)$ is increased by 1, if possible (otherwise there is no transition). The left moves ($h = -1$ are dual).

In addition, upon moving between segments, $\mathcal{V}$ may nondeterministically record the current state and segments in `loop_statea` and `loop_seg_move` respectively. Note that from then on, `c_freeze_1` no longer changes. If those are already recorded then it may check whether the current state and segments match the recorded ones, in which case we move to the 2-NFA Component (which also freezes `c_freeze_2`).

If at any point we have `cur_state` $= q_{\text{acc}}$, we move to the Finalizing Component.

**2-NFA Component.** A state contains the following information:

- `cur_state` $\in Q$ – the current state of $\mathcal{A}$.
- `cur_seg` $\in \{0, \ldots, k+1\}$ – the current segment, with 0 and $k+1$ representing $\vdash$ and $\dashv$, respectively.
- `num_moves`: $\{(i, i+1), (i+1, i) \mid i \in \{0, \ldots, k\}\} \to \{0, \ldots, |Q|+1\}$ – how many times did we move from segment $i$ to segment $i+1$ (bounded by $|Q|+1$).
- `loop_seg_move` $\in \{(i, i+1), (i+1, i) \mid i \in \{0, \ldots, k\}\} \cup \{\bot\}$ – the border between segments where we guess a loop occurs, or $\bot$ if this is not guessed yet.
- `loop_state` $\in Q \cup \{\bot\}$ – the state where we guess a loop occurs, or $\bot$ if this is not guessed yet.

The transitions are similar to those of the Simulation Component, with the exception that `c`, `c_freeze_1`, `c_freeze_2` are no longer changed (i.e., the counter of $\mathcal{A}$ is no longer tracked). Note that upon reaching the 2-NFA Component, the function `num_moves` is reset to 0 to begin a fresh count.

**Finalizing Component.** This component consists of two states: $s_{\text{check_loop}}$ and $s_{\text{fin}}$. $s_{\text{check_loop}}$ has a self loop with operation `c_freeze_1` $\leftarrow -1$, `c_freeze_2` $\leftarrow -1$, and a transition to $s_{\text{fin}}$ with operation `c_freeze_2` $\leftarrow -1$. Lastly, $s_{\text{fin}}$ has a self loop allowing to decrease each counter separately, except for `c_freeze_1`.

We prove the correctness of the construction in Appendix A.

▶ **Lemma 17.** *$L(\mathcal{A}) \neq \emptyset$ iff $(s_{fin}, \mathbf{0})$ is reachable from $(s_{init}, \mathbf{0})$ in $\mathcal{V}$.*

**Proof.** It is not hard to see that by construction, the states of the Simulation Component of $\mathcal{V}$ correctly track the word guessed by $\mathcal{V}$ in the Initialization Component provided no more than $|Q|$ segment moves occur. Thus, in order to prove correctness, it is enough to prove that if $\mathcal{A}$ accepts some word $w$, then it also accepts $w$ with at most $|Q|$ moves between segments $i$ and $i+1$ for all $i$. Unfortunately, this is not always the case. However, note that if more than $|Q|$ moves between segments $i$ and $i+1$ occur, then there must be a state $q$ that is repeated in two of those moves. If the counter effect of this loop is nonpositive, then the loop can be cut to obtain a "better" run (i.e., one whose counter values are not lower). Otherwise, if the counter effect of the loop is positive, then it can be repeated to make the counter of $\mathcal{A}$ arbitrarily high. Then, $\mathcal{A}$ accepts the word iff the 2-NFA obtained from $\mathcal{A}$ by discarding the counter effects is accepted from the loop state $q$ and the position of the head at the end of segment $i$. Moreover, such an accepting run can be assumed not to loop, since otherwise a shorter accepting run can be found. Thus, In the latter case, the 2-NFA component tracks an accepting run, if one exists.

Finally, note that the finalization component can reach $s_{\text{fin}}$ with `c_freeze_1` $= 0$ only if `c_freeze_1` $<$ `c_freeze_2`, i.e., the loop taken was indeed strictly positive. ◀

## B Proof of Theorem 6

We present the detailed construction of $\mathcal{V}$ from $\mathcal{A}$, starting with the counters of $\mathcal{V}$ and their intuitive roles.

- `f[1]`, ..., `f[|Q|+1]` – the counter value before each forward sweep.
- `b[1]`, ..., `b[|Q|+1]` – the counter value before each backward sweep.
- `f_loop_1` – the counter before the first forward sweep that starts with a loop state $q$.
- `f_loop_2` – the counter before the second forward sweep that starts with a loop state $q$.

We now turn to describe the states and transitions of $\mathcal{V}$, split into three components.

**Initialization Component.** This component starts in state $s_{\text{init}}$ and has other states containing the following information:

- qf[1], ..., qf[|Q|+1] $\in Q \cup \{\bot\}$ – the state *beginning* each forward sweep, or $\bot$ if the sweep is not taken.
- qb[1], ..., qb[|Q|] $\in Q \cup \{\bot\}$ – the state *ending* each backward sweep, or $\bot$ if the sweep is not taken.
- q_loop – a state that appears twice in the forward sweeps.

The transitions from $s_{\text{init}}$ guess the components in the next state, with the following restrictions:

- if $\bot$ appears at some entry, then all later entries are also $\bot$ (i.e., once a certain sweep is not taken, not further sweeps are taken).
- the first forward state is forced to be $q_{\text{init}}$ and if q_loop $= \bot$ then the last forward state that is not $\bot$ is $q_{\text{acc}}$.
- qb[i] $=$ qf[i+1] for all $i \leq |Q|$, i.e., the backward sweep ends with the same state the next forward sweep starts with.
- The field q_loop is guessed in a consistent way, i.e., it must be the first state that appears twice in the qf[i] states.

Each initialization state has self loops to charge the counters with the following operations: for every $i \in 1, \ldots, |Q|$, we have a loop with operations b[i] $\leftarrow +1$, f[i+1] $\leftarrow +1$, corresponding to the fact that the counter with which the $i$-th backward sweep ends is the same as the counter with which the $i+1$-th forward sweep starts. In addition, if q_loop $=$ qf[i] $=$ qf[j] for some $i < j$, then f_loop_1 $\leftarrow +1$ is applied together with f[i] $\leftarrow +1$ and f_loop_2 $\leftarrow +1$ together with f[j] $\leftarrow +1$. Thus enforcing the semantics mentioned above.

The initialization states can nondeterministically move to the Simulation Component.

**Simulation Component.** The initialization component keeps track of the forward and backward runs of $\mathcal{A}$, as well as acceptance in the DFA equivalent to $\mathcal{A}$ from the loop state q_loop as follows. Each state in this component carries the same information as the initialization component, but updates it using the transition function.

First, we obtain $\mathcal{A}^R = \langle Q, \Sigma, \Delta^R, q_{\text{acc}}, q_{\text{init}} \rangle$ by defining $\Delta^R = \{(q', \sigma, -e, -h, q) \mid (q, \sigma, e, h, q') \in \Delta,\ \sigma \in \Sigma\}$ (observe that we omit transitions on $\vdash, \dashv$, this is explained below). Thus, $\mathcal{A}^R$ essentially reverses the runs of $\mathcal{A}$. Furthermore, for each state $q \in Q$ we obtain from $\mathcal{A}$ a DFA $\mathcal{D}_q$ by discarding the counter effects of $\mathcal{A}$, setting the initial state to $q$ and converting the resulting 2-NFA to a DFA [31, 30, 32] of single-exponential size.

The main behavior in this component is as follows. From state $s$ of $\mathcal{V}$, we guess a letter $\sigma \in \Sigma_{\vdash,\dashv}$ (the guesses of $\vdash$ and $\dashv$ have special status, see below). The transition then guesses, for each state qf[i] in $s$, a transition on $\sigma$, and updates the corresponding f[i] counter with the transition effect. Dually, for each [$qb[i]$] we guess a transition on $\sigma$ in $\mathcal{A}^R$ and update b[i] accordingly. Finally, we update *q_loop* using the DFA $\mathcal{D}_q$ with the transition on $\sigma$.

To treat the end-markers, we note that reading an end-marker is joint between the forward and backward runs, and we therefore only want to simulate each end-marker in one of them, and we choose the forward runs (which is why we omit these transitions from $\mathcal{A}^R$). We force the first letter guessed to be $\vdash$, and we only update the forward elements (the qf[i] states and f[i] counters). Upon guessing $\dashv$, we again only update the forward elements, and we transition to the Finalizing Component.

**Finalizing Component.** In order to verify that the sweeps are concatenated correctly, we proceed as follows. In state $s_{\text{check_sweeps}}$ we first check that $\texttt{qf[i]} = \texttt{qb[i]}$, i.e., that the forward and backward sweeps meet at the same state. We then have loops with operation $\texttt{f[i]} \leftarrow -1, \texttt{b[i]} \leftarrow -1$ for all $1 \le i \le |Q|$. Note that such transitions can be taken to reach $\mathbf{0}$ iff the counters are the same, implying that the forward and backward sweeps meet with the same counter value.

The above checks that the forward and backward sweeps form a valid run of $\mathcal{A}$ on the guessed word. If the last forward state is $q_{\text{acc}}$, this ensures the word is accepted. It remains to check acceptance in the case of a loop. To this end, if $\texttt{q_loop} \neq \bot$, we check that the state reached in $\texttt{q_loop}$ is an accepting state of $\mathcal{D}_q$. However, we also need to check that the loop taken is indeed positive. To achieve this we move to a state with operation $\texttt{f_loop_1} \leftarrow -1, \texttt{f_loop_2} \leftarrow -1$, and then a final transition with only the operation $\texttt{f_loop_2} \leftarrow -1$ to $s_{\text{fin}}$. Then, reaching $\mathbf{0}$ ensures that the loop was indeed positive, as the counter increased between the first and second visits to the loop states.

The correctness of the construction follows from the observation that if $w \in L(\mathcal{A})$, then either $w$ accepted within $|Q|$ forward sweeps, or a positive loop is reached, and from this loop there is a run to an accepting state. In this case, the loop can be arbitrarily pumped so that the run to the accepting state can be taken in the corresponding DFA.

Finally, since the size of $\mathcal{D}_q$ is single-exponential in the size of $\mathcal{A}$ [31, 30, 32], the construction is single-exponential. ◀

## C Proof of Lemma 15

We present the complete proof of Lemma 15 in this section.

Consider a 2-DOCN $\mathcal{A} = \langle Q, \Sigma, \Delta, q_{\text{init}}, q_{\text{acc}} \rangle$. For a word $w$, we denote by $\Upsilon$ the set of initial runs of $\mathcal{A}$ on $w$. Note that when $\mathcal{A}$ is deterministic, the runs in $\Upsilon$ are ordered by the prefix relation $\prec$, so we refer to $\Upsilon$ as the set of *run prefixes*. In addition, $\Upsilon$ may be infinite, in case $\mathcal{A}$ has unboundedly long run prefixes (and in particular, since $\mathcal{A}$ is deterministic, it does not accept $w$).

Denote by $M = |Q|$ the number of states in $\mathcal{A}$. Our goal is to devise a way to pump certain infixes of a word, so that we can reason about the resulting run of $\mathcal{A}$ on it. Naturally, the challenge lies in synchronizing the behaviors of forward and backward traversals of the run.

We set the bound $K$ prescribed in Lemma 15 to be $K = M + (M+1)^{2(M+1)} + 1 + M$, and we think of $a^K$ as partitioned to three parts: the *outer left a's* $a^M$, the *inner a's* $a^{(M+1)^{2(M+1)}+1}$ and the *outer right a's* $a^M$. In the following, we consider segments of the run of $\mathcal{A}$ on a word $w = xa^Ky$ with $x, y \in \Sigma^*$. We start by observing that if the run of $\mathcal{A}$ never moves the head beyond $x$, then the lemma is trivial. We therefore assume that the head does go beyond $x$, and we start all our reasoning assuming that this prefix of the run was already taken, i.e., that the head now lies on the last letter of $x$.

▶ Remark 18 (Drawing Convention). Our convention for depicting run types uses 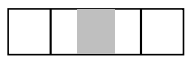 to depict the word $xa^Ky$, with the grayed area representing the inner $a$'s, and the vertical lines marking the end of $x$ and the beginning of $y$. Runs are depicted as going from top to bottom upon head reversal, and hence e.g., 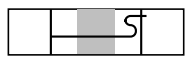 depicts runs going left, and 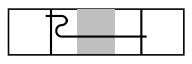 depicts runs going right. We remark that the drawings serve no formal role, only a convention to assist the naming of run types.

▶ **Definition 19.** *Consider a run $\rho = (q_1, c_1, p_1), \ldots, (q_n, c_n, p_n)$. We say that $\rho$ is:*

- *a* left loop*, denoted $\rho \in$ , if $p_1 = p_n = 0$ and for every $1 \le i \le n$ we have $p_i \le K$.*
- *a* right loop*, denoted $\rho \in$ , if $p_1 = p_n = K+1$ and for every $1 \le i \le n$ we have $p_i > 0$.*
- *a* left sink*, denoted $\rho \in$ , if $p_1 = 0$, for every $1 \le i \le n$ we have $p_i \le K$ and there exist $1 \le j < n$ such that $q_j = q_n$ and $p_j = p_n$.*
- *a* right sink*, denoted $\rho \in$ , if $p_1 = K+1$, for every $1 \le i \le n$ we have $p_i > 0$ and there exist $1 \le j < n$ such that $q_j = q_n$ and $p_j = p_n$.*
- *a* right crossing*, denoted $\rho \in$ , if $p_1 = 0$, $p_n = K+1$, for every $1 < i < n$ we have $0 < p_i < K+1$.*
- *a* left crossing*, denoted $\rho \in$ , if $p_1 = K+1$, $p_n = 0$, for every $1 < i < n$ we have $0 < p_i < K+1$.*
- *a* round trip *if $\rho = l \cdot c_r \cdot r \cdot c_l$ with*[3] *$l \in$ , $c_r \in$ , $r \in$ and $c_l \in$ .*

Our first claim is that if the run of $\mathcal{A}$ moves from $x$ and crosses from the outer left $a$'s to the inner $a$'s, then the run must complete the traversal of $a$'s and cross over to $y$ before seeing $x$ again (and dually for right-to-left runs). Intuitively, this is because $\mathcal{A}$ is deterministic and therefore cannot read many $a$'s before entering a right-traversing cycle.

▶ **Lemma 20.** *Consider a run $\rho = (q_1, c_1, p_1) \cdots (q_n, c_n, p_n)$. If $\rho \in$ (left loop) or $\rho \in$ (left sink), then $p_i \le |x| + M$ for all $1 \le i \le n$. Similarly, if $\rho \in$ (right loop) or $\rho \in$ (right sink), then $p_i \ge |x| + K + 1 - M$ for all $1 \le i \le n$.*

**Proof.** Consider the case where $\rho \in$ , and assume by way of contradiction that $\rho$ crosses to position $|x| + (M+1)$. Then the infix of $\rho$ before the first time it reaches position $|x| + (M+1)$ contains a sequence of at least $M+1$ configurations within the left-outer $a$, upon which the head moves to the right. Thus, $\rho$ has a cycle from some state $q$ upon reading $a$'s that moves the head some $j$ positions to the right. This cycle is then repeatedly taken until a letter other than $a$ is encountered, which only occurs on $y \dashv$. Thus, $\rho$ must visit position $|x| + K$ before reaching position 0 again, in contradiction to it being a left-loop.

The analysis for the remaining three types of runs is similar. ◀

We now turn to identify forms of runs of $\mathcal{A}$ on $xa^K y$, based on their decomposition to round trips (see Definition 19). Recall that while $\mathcal{A}$ is deterministic, it may have many (possibly infinitely many) runs on $xa^K y$, ordered by the prefix relation.

▶ **Definition 21.** *Consider a run $\rho$ of $\mathcal{A}$ on $xa^K y$. We say that $\rho$ has:*

1. Form 1: Many round trips *if $\rho = r_1 \cdot r_2, \cdots r_{M+1}$ for round trips $r_1, \ldots, r_{M+1}$.*
2. Form 2: Few round trips and a sink *if $\rho = r_1 \cdot r_2, \cdots r_m \cdot I$ for round trips $r_1, \ldots, r_m$ where $m \le M$ and $I$ is either (left sink) or $I = l \cdot c_r \cdot d$ with $l \in$ , $c_r \in$ and $d \in$ .*
3. Form 3: Few round trips cut short *if $\rho = r_1 \cdot r_2, \cdots r_m \cdot C$ for round trips $r_1, \ldots, r_m$ where $m \le M$ and $C$ is a prefix of a round trip.*

▶ **Lemma 22.** *Let $\Upsilon$ be the set of run-prefixes of $\mathcal{A}$ on $xa^K y$, then either there is a run $\rho \in \Upsilon$ of Form 1 or of Form 2, or $\Upsilon$ is finite and the maximal-length run in $\Upsilon$ is of Form 3.*

[3] We often concatenate runs $\alpha \cdot \beta$ where $\alpha$ ends with the same configuration $\beta$ starts with. In such cases, the concatenation means that we omit the overlapping configuration.

**Proof.** Recall that $\mathcal{A}$ does not accept $xa^K y$, and therefore there is no accepting run. The main observation required in the proof is that by Lemma 20, if a left loop is longer than $M(|x|+M)+1$, then it has a prefix that is a left sink. Indeed, a left loop must stay within the outer-left $a^M$, and uses at most $M$ states. Similarly, a right loop longer than $M(|y|+M)+1$ has a prefix that is a right sink. Still by Lemma 20, a run that starts at position $|x|$ (resp. $K+1$) and reaches the inner $a$'s is a right crossing (resp. left crossing).

Therefore, runs from position 0 start as a (possibly empty) left loop, followed by a right crossing, then a right loop, then a. This repeats until either the run cannot continue due to the counter becoming negative, or due to one of the loops becoming a sink.

The lemma now readily follows: if $\Upsilon$ has a long enough run of Form 1, we are done. Otherwise all runs in $\Upsilon$ have at most $M$ round trips. Then, either $\Upsilon$ has a run that ends with a left or right sink, or $\Upsilon$ is finite and the maximal run ends due to the counter becoming negative (i.e., ends with a prefix of a round trip). ◀

Having established the form of runs, we now turn to our main pumping argument. To this end, we identify within the inner $a$'s two positions that form a "cycle" with respect to left and right crossings.

Let $c$ be a left or right crossing. We say that two positions $p_1, p_2$ are *$c$-matching positions* if $|x|+M+1 \leq p_1 < p_2 \leq |x|+K-M$ (i.e., these two positions are in the inner $a$'s.) and the first time positions $p_1, p_2$ occur in $c$ – they occur with the same state.

We extend the definition to prefixes of $c$: for a prefix $c' \prec c$, we say that $|x|+M+1 \leq p_1 < p_2 \leq |x|+K-M$ are $c'$-matching positions if either both positions occur in $c'$ and appear for the first time with the same state, or both positions do not occur in $c'$.

Next, we lift the definition to concatenations of runs: for a run $\rho$ that is a concatenation of the types in Definition 19 and prefixes thereof, we say that $|x|+M+1 \leq p_1 < p_2 \leq |x|+K-M$ are matching positions if they are matching for every crossing (or prefix of a crossing) in $\rho$. In particular, for a round trip $R = l \cdot c_r \cdot r \cdot c_l$, positions $p_1, p_2$ are matching if they are matching for both $c_r$ and $c_l$.

▶ **Lemma 23.** *Let $\rho$ be a run of $\mathcal{A}$ on $xa^K y$ in one of the Forms of Definition 21, and let $|x|+M+1 \leq p_1 < p_2 \leq |x|+K-M$ be matching positions for $\rho$. Let $w^+ = xa^{K+(p_2-p_1)}y$ and $w^- = xa^{K-(p_2-p_1)}y$, then there is are runs $\rho^+$ and $\rho^-$ of $\mathcal{A}$ on $w^+$ and $w^-$, respectively, such that each of them either ends in the same state as $\rho$, or ends due to the counter becoming negative.*

**Proof.** We prove the claim for a single round trip $R = l \cdot c_r \cdot r \cdot c_l$ where $l \in$, $c_r \in$, $r \in$ and $c_l \in$. The claim follows inductively for a concatenation of round trips, and is analogous for the suffixes of Forms 2 and 3 in Definition 21.

We start with $w^+ = xa^{K+(p_2-p_1)}y$. Denote $R = l \cdot c_{r1}c_{r2}c_{r3} \cdot r \cdot c_{l1}c_{l2}c_{l3}$, such that $c_r$ is split into three parts – before the first occurrence of $p_1$, between the first occurrence of $p_1$ and the first occurrence of $p_2$, and after the first occurrence of $p_2$. Similarly, $c_l$ is split so that $c_{l1}$ is before the first occurrence of $p_2$ (since this is a left crossing), $c_{l2}$ is between $p_2$ and the first occurrence of $p_1$, and $c_{l3}$ is the rest.

For a run infix $\beta$, we denote by $\beta^{+t,+t'}$ the run obtained from $\beta$ by offsetting each configuration $(q,c,p)$ to $(q,c+t,p+t')$. We now consider the run

$$l \cdot c_{r1} \cdot c_{r2} \cdot (c_{r2} \cdot c_{r3} \cdot r \cdot c_{l1} \cdot c_{l2})^{+\mathrm{eff}(c_{r2}),+(p_2-p_1)} \cdot (c_{l2} \cdot c_{l3})^{+\mathrm{eff}(c_{r2})+\mathrm{eff}(c_{l2}),+0}$$

Observe that up to the counter updates, this is a run of $\mathcal{A}$ on $w^+$. Indeed, since $p_1$ and $p_2$ are matching positions, then we can take in $p_2$ the same transitions of $c_{r2}$, again reaching the same initial state of $c_{r3}$. Moreover, since $p_1, p_2$ are in the inner $a$'s, Lemma 20 guarantees that within the pumped infix $c_{r2}$, the head does not reach positions $|x|$ nor $|x|+K+1+(p2-p1)$ (otherwise this would also happen before pumping). Similar arguments apply for the left crossing. Thus, either this is a run on $w^+$ ending in the same state and position as $\rho$, or a prefix of this run ends due to the counter becoming negative.

Similarly, for $w^-$ we obtain the run

$$l \cdot c_{r1} \cdot (c_{r3} \cdot r \cdot c_{l1})^{-\mathrm{eff}(c_{r2}),-(p_2-p_1)} \cdot c_{l3}^{-\mathrm{eff}(c_{r2})-\mathrm{eff}(c_{l2}),+0}$$

and again, similar consideration to Lemma 20 show that upon reading $c_{r3}$ the run does not reach position 0 before proceeding to $r$. ◀

We are now ready to complete the proof of Lemma 15. Recall that our goal is to find $K' \neq K$ such that $xa^{K'}y \notin L(\mathcal{A})$.

**Proof of Lemma 15.** We follow the run classification given by Definition 21 and Lemma 22, and split into cases according to the form of the run. Let $\Upsilon$ be the set of run-prefixes of $\mathcal{A}$ on $xa^K y$.

We start with a general counting argument for all the forms. Consider a run $\rho$ in one of the forms of Definition 21, i.e., a concatenation of either $M+1$ round trips, or at most $M$ round trips and a suffix containing at most two crossings.

Recall that the inner $a$'s sequence is of length $(M+1)^{2(M+1)}+1$. We associate with each position $|x|+M < p < |x|+K-M$ a function $f_p \colon \{1,\ldots,2(M+1)\} \to Q \cup \{\bot\}$ such that $f_p(i)$ is the first state reached in position $p$ on the $i$-th crossing in $\rho$, or $\bot$ if this position is not reached in the $i$-th crossing (this can happen only in the suffix in Form 3, or if there are fewer than $M+1$ round trips in Forms 2 or 3). Since there are at most $(M+1)^{2(M+1)}$ such functions, it follows that within the inner $a$'s, there are two positions $|x|+M < p_1 < p_2 < |x|+K-M$ for which $f_{p_1} = f_{p_2}$. Thus, $p_1$ and $p_2$ are matching positions for $\rho$.

We now split to cases, starting with the most involved Form.

1. (*Form 3*) If $\Upsilon$ is finite and its maximal run is $\rho = r_1 \cdots r_m C$, for round trips $r_1, \ldots r_m$, and $C$ a prefix of a round trip, let $p_1 < p_2$ be $\rho$-matching positions as above. We consider the counter effects $e_r^i$ and $e_l^i$ of the run between $p_1$ and $p_2$ and between $p_2$ and $p_1$ on the right and left crossing in the round trip $r_i$.
   If $e_r^1 < 0$, then by considering the word $xa^{K+B(p_2-p_1)}y$ for large enough $B$, the run of $\mathcal{A}$ reaches a negative counter value, and is not accepted. Thus, setting $K' = K+B(p_2-p_1)$, we have $K' \neq K$ such that $xa^{K'}y \notin L(\mathcal{A})$ and we are done. Otherwise, if $e_r^1 \geq 0$ and $e_r^1 + e_l^1 < 0$, then again the run on the word $xa^{K+B(p_2-p_1)}y$ reaches a negative counter for large enough $B$. Indeed, the counter accumulated in $e_r^1$ cannot compensate for the decrease of $e_l^1$. In a similar fashion, if at any point $e_r^1 + e_l^1 + \ldots + e_r^i < 0$ or $e_r^1 + e_l^1 + \ldots + e_r^i + e_l^i < 0$, then by taking large enough $B$ we obtain a word not accepted by $\mathcal{A}$.
   The remaining case is that all the partial sums are non-negative (and in particular their entire sum). Recall that $\rho$ is the maximal-length run in $\Upsilon$. We now consider the word $w^- = xa^{K-(p_2-p_1)}y$, then by Lemma 23 we have that either the run $\rho^-$ of $\mathcal{A}$ on $w^-$ reaches a negative counter and therefore does not accept, or it ends at the same state as $\rho$. In the latter case, observe that the counter value at the end of $\rho^-$ is not larger than that in $\rho$, since $\rho^-$ was obtained by removing an infix with overall non-negative effect, namely the sum of the effects of the cut cycles. Thus, $\rho^-$ is also maximal, and therefore $\mathcal{A}$ does not accept $w^-$, so we are done.

2. (*Form 1*) If $\Upsilon$ contains a run $\rho = r_1 r_2 r_3 \cdots r_{M+1}$ for round trips $r_1, \ldots r_{M+1}$, let $p_1 < p_2$ be $\rho$-matching positions as above. By Lemma 23 we get that either $w^+ = xa^{K+(p_2-p_1)}y$ has a run that ends in the same (non accepting) state as $\rho$, or $w^+$ is not accepted due to the counter becoming negative. In the latter case, we found $K' \neq K$ such that $xa^{K'}y \notin L(\mathcal{A})$ and we are done. In the former case, since the run on $w^+$ has $M+1$ round trips, then there is a state $q \in Q$ that is visited twice at position 0 at the beginning of two distinct round trips. This implies that $\mathcal{A}$ repeats a cycle on $w^+$, and therefore either never halts, or at some point makes the counter negative. In either case, $w^+ \notin L(\mathcal{A})$ and again we are done.
3. (*Form 2*) If $\Upsilon$ contains a run $\rho = r_1 \cdots r_m I$, for round trips $r_1, \ldots r_m$, and $I$ ending with a left sink 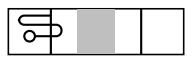 or right sink 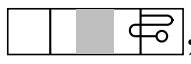, then again we consider the $\rho$-matching positions $p_1, p_2$ and $w^+ = xa^{K+(p_2-p_1)}y$. Again by Lemma 23, either the run on $w^+$ also reaches a sink, or it reaches a negative counter, and anyway does not accept, so we are done. ◀